\documentclass{article}
\usepackage{spconf}
\usepackage{tabu}

\usepackage[]{siunitx}

\usepackage{cite}
\usepackage{graphicx}
\usepackage[cmex10]{amsmath}
\usepackage{multirow}
\usepackage{makecell}
\usepackage{diagbox}

\usepackage{newtxtext,newtxmath}
\usepackage{bm}
\usepackage{psfrag}
\usepackage{color}
\usepackage{colortbl}

\title{VarArray: Array-Geometry-Agnostic Continuous Speech Separation}
\makeatletter
\def\@name{\textit{Takuya Yoshioka, Xiaofei Wang, Dongmei Wang, Min Tang, Zirun Zhu, Zhuo Chen, Naoyuki Kanda}\\
}
\makeatother
\address{Microsoft, One Microsoft Way, Redmond, WA, USA}

\begin{document}
\ninept

\maketitle

\begin{abstract}
Continuous speech separation using a microphone array was shown to be promising in dealing with the speech overlap problem in natural conversation transcription. This paper proposes VarArray, an array-geometry-agnostic speech separation neural network model. The proposed model is applicable to any number of microphones without retraining while leveraging the nonlinear correlation between the input channels. The proposed method adapts different elements that were proposed before separately, including transform-average-concatenate, conformer speech separation, and inter-channel phase differences, and combines them in an efficient and cohesive way. Large-scale evaluation was performed with two real meeting transcription tasks by using a fully developed transcription system requiring no prior knowledge such as reference segmentations, which allowed us to measure the impact that the continuous speech separation system could have in realistic settings. The proposed model outperformed a previous approach to array-geometry-agnostic modeling for all of the geometry configurations considered, achieving asclite-based speaker-agnostic word error rates of 17.5\% and 20.4\% for the AMI development and evaluation sets, respectively, in the end-to-end setting using no ground-truth segmentations. 
\end{abstract}
\section{Introduction}
\label{sec:Introduction}

The last decade has witnessed transformational progress in 
automatic speech recognition (ASR) technology driven by deep learning approaches. 
Modern ASR systems can accurately transcribe pre-segmented utterances recorded in acoustically moderate environments, as demonstrated in many ASR benchmarks such as LibriSpeech~\cite{zhang2020pushing} and Switchboard~\cite{Tuske21}, and are widely used in our daily lives. 
Yet, there are many challenges that must be overcome for the ASR technology to become usable more broadly~\cite{szymanski2020wer,delrio21_interspeech}. 
One of the unsolved challenges is transcribing unsegmented natural conversations, where multiple people talk over each other in a spontaneous and thus unpredictable way. 

Continuous speech separation (CSS) was proposed to handle the speech overlaps in natural human-to-human conversation transcription~\cite{Yoshioka18b,Yoshioka19d,LibriCSS}. 
With this approach, continuously captured long-form conversational signals are split into multiple overlap-free signals. 
Since each of the output signal does not contain the speech overlaps, they can be processed with conventional ASR systems. 
A microphone array-based approach has been particularly successful and was applied to real meeting recordings~\cite{Yoshioka19d} while most previous studies used artificially created meeting-like data. 


This paper sheds light on the generalizability of the multi-channel speech separation models.
Two major approaches exist for the multi-channel neural network modeling for speech separation. One approach uses multi-channel features confined to a specific microphone array geometry~\cite{Yoshioka18,Wang21-complexspectralmapping}. 
While it can potentially utilize the full spatial information, the trained model can be used only for the  microphone array that it is trained for. 
Another approach is to apply the same speech separation model to each input channel and merge the outputs from the individual channels by calculating their median or average~\cite{Chang20-e2e,Zhang21-stability}. While being applicable to any microphone array devices, this approach does not take advantage of the nonlinear correlation between the channels. Also, the computational cost becomes too large to deploy in practice when the number of channels is large. 

This problem is addressed by VarArray, our proposed array geometry-agnostic speech separation model.
The input is a set of magnitude and inter-channel phase difference (IPD) features, organized in a way that is invariant to the microphone permutation. 
The model consists of conformer blocks~\cite{gulati20_interspeech,Chen21-conformersep} interleaved with transform-average-concatenate (TAC) layers~\cite{Luo20-TAC} to exploit the spatio-temporal patterns exhibited in the input. 
TAC is a cross-channel layer model that can cope with any number of channels in a permutation-invariant fashion. 
To reduce the growth rate of the model evaluation cost with respect to 
the number of microphones, the multiple feature streams are merged early in the network.

Large-scale evaluation was carried out by using real meeting recordings. 
Two meeting corpora were used to evaluate the performance sensitivity to the array geometry: AMI~\cite{AMICorpus} and Microsoft internal meetings~\cite{Yoshioka19d}. 
Our meeting transcription system achieved state-of-the-art results in AMI under the condition of no ground-truth segmentations being used. 
Results of fine-tuning the speech separation model to AMI with an ASR-based loss function are also presented, demonstrating its impact on the full-blown system.
Note that, 
while the array-geometry-agnostic modeling was studied before in related areas, such as dereverberation~\cite{yemini21_interspeech}, 
noise reduction~\cite{zhang21e_interspeech}, and ad hoc microphone arrays~\cite{wang20c_interspeech}, 
this paper is the first to combine different pieces including the IPD features, conformer speech separation, and TAC in a cohesive and efficient way and presents thorough evaluation results for the real meeting tasks.




\section{Continuous Speech Separation}

\subsection{General architecture}
\label{sec: css_architecture}
\vspace{-.3em}

CSS is a front-end-based approach to the speech overlap problem in natural human-to-human conversation transcription. 
Given long-form audio signals captured by a microphone array, 
the goal of CSS is to generate $K$ signals that have the same length as the input in such a way that each of the $K$ signals does not contain overlapped utterances inside while the sum of these $K$ signals retains the spoken content present in the input signals in its entirety. 
It is tempting to try to estimate the clean speech signal of each speaker, which means $K$ is equal to the total number of speakers. However, this approach is extremely difficult for streaming processing because it requires the front-end to perform speaker diarization in addition to enhancing the speech.  
A more practical approach, which is increasingly being adopted, is to address only the speech overlap problem~\cite{Yoshioka19,LibriCSS,Wang21-complexspectralmapping,GraphPIT}. 
In meetings, the maximum number of simultaneously talking speakers is very limited. In fact, we can assume only two or fewer speakers to be active for the majority of the meeting time~\cite{Cetin06}. 
This means that only two output signals are sufficient in practice (i.e., $K=2$). 
If the audio segment being processed has two overlapping utterances, we separate each utterance and send the obtained signals to different output channels. 
When the current segment consists of only one speaker, we just have to enhance the speech quality and route the processed signal to one of the output channels. 
The unused output channel can be filled by zero to make the two output signals synchronous. 
As the output signals are overlap-free, 
they can be directly fed to a conventional ASR system.

A sliding window-based approach is often adopted to enable CSS. 
At each window position, 
a speech separation model trained with utterance-level permutation invariant training (uPIT)~\cite{uPIT} is applied to the windowed input signal to generate $K$ separated signals. 
The order of the output signals is determined to maintain the consistency with the separated signals obtained at the previous window position.
Specifically, we choose the output permutation that provides 
the smallest mean squared error between the pairs of the separated signals obtained at the current and previous window positions. The errors are calculated for the overlapping frames of the two windows. 
This ``stitching processing''  can be done by using the overlapped segment between the two adjacent windows.
In the following, we focus our attention on the local speech separation and uPIT as it constitutes the central piece of the sliding window-based approach. 

\subsection{Local speech separation}
\label{sec: local_separation}
\vspace{-.3em}

The local speech separation process can be described as follows. 
Let $X_{mft}$ denote the short time Fourier transform (STFT) coefficient of the audio signal observed by the $m$th microphone, where $f$ and $t$ represent the frequency and time indices, respectively. 
Feature vector $\bm{y}_t$ is computed for each time frame, and the sequence of the feature vectors is fed to a speech separation neural network model. 
The model produces time-frequency (TF) masks for predicting 
the two clean speech signals (recall that we assume $K=2$) and the background noise. 
The masks are used to perform minimum variance distortionless response (MVDR) beamforming to estimate the clean signals. 
It was empirically found that sparcifying the TF masks before the MVDR computation improved the ASR accuracy. 
In our experiments, this was performed by retaining only the most dominant sound source for each TF bin. 
Gain adjustment of \cite{Yoshioka18} was also applied after beamforming to counteract MVDR's inability of generating complete silence for non-overlapped signals.

During training, reference signals are available for the clean speech and the noise. For training samples with one active speaker, 
one of the reference signals is a zero sequence. 
The uPIT loss~\cite{uPIT} is employed to cope with the arbitrariness of the order of the two speech signals. 

The feature vector, $\bm{y}_t$, can be the concatenation of the STFT coefficients of all the microphones~\cite{Wang21-complexspectralmapping} or a stack of IPDs~\cite{Yoshioka18}, just to mention a few examples. 
Various speech separation models can be applied atop these concatenated features, including conformer networks~\cite{Chen21-conformersep} and dual-path networks~\cite{DPRNN}. 
However, using the concatenated features make the system dependent on a particular microphone array and usable only with the audio device which it is trained for. 
Our goal is to create a speech separation model that can work with any microphone arrays which performs as well as or better than the array geometry-dependent ones. 



\section{VarArray}

VarArray, our proposed speech separation model, can efficiently utilize the spatial and temporal information from any number of microphone inputs for accurate TF mask estimation. It also yields the same output regardless of the microphone permutation. 
There are two key ingredients: one concerns the feature design; the other relates to the model structure. They are described in 
Sections 3.1 and 3.2, respectively. 
In the following, $\mathbb{M}$, $\mathbb{F}$, and $\mathbb{T}$ denote 
the index sets of the microphones, frequency bins, and time frames, respectively.

\subsection{Feature set}
\vspace{-.3em}

With VarArray, we use a variable-size feature set, $\{\bm{z}_{mt}\}_{m \in  \mathbb{M}}$, for each time frame $t$ instead of stacking all the features obtained from different microphones in a particular order. 
This prevents the input features from being tied to a specific microphone array with a predefined channel indexing system. 
The simplest way is to use the STFT coefficients of the $m$th microphone as $\bm{z}_{mt}$. 
Below, we propose a normalized IPD-based feature set, which outperformed the STFT-based method in our preliminary experiment.  


With the proposed feature set, $\bm{z}_{mt}$ is defined as the concatenation of the magnitude of an average spectrum and the channel-wise IPD with respect to the average spectrum. That is, we have $\bm{z}_{mt} = [ \bm{z}^{\text{mag}}_{mt}, \bm{z}^{\text{ipd}}_{mt} ]$,
where 
$
\bm{z}^{\text{mag}}_{mt} = ( | \Bar{X}_{ft} |^2 )_{f \in \mathbb{F}} \text{~and~}
\bm{z}^{\text{ipd}}_{mt} = \angle ( X_{mft} / \Bar{X}_{ft} )_{f \in \mathbb{F}}
$
with $\Bar{X}_{ft}$ being the average of the STFT coefficients over the microphones, $\{X_{mft}\}_{m \in \mathbb{M}}$.
Both the magnitude and IPD features are normalized at the window level to reduce random fluctuations, whose effectiveness was experimentally shown in \cite{Yoshioka18}.

\subsection{Speech separation model}
\vspace{-.3em}

Now, we turn our attention to the modeling side of this work. 
The objective is to construct a speech separation model that receives 
the feature sequence $( \bm{z}_{mt} )_{t \in \mathbb{T}}$ from all the input channels and generates 
TF mask $M_{sft}$ over the input segment, where $s$ denotes the source index and takes a value in $\{0, \cdots, 3\}$.
The first two sources (i.e., $s \in \{0, 1\}$) correspond to the two speakers. 
The other two sources are used to capture the stationary ($s = 2$) and transient ($s = 3$) noise. This is inspired by the SSN architecture proposed in \cite{Yoshioka18b}, where SSN stands for ``speech, speech, and noise''. 
It is required that 
the output be invariant to the input channel permutation and that 
the model be applicable to any $M$ value. 

Several previous studies applied a split-apply-combine (SAC) approach. 
That is, the same speech separation model is applied to each input channel independently. Then, the TF masks obtained from all the $M$ channels are combined, for example, by taking their average~\cite{Chang20-e2e,Zhang21-stability}. 
The output mask order may be realigned across the $M$ channels before the average computation~\cite{Wang19-spectralspatial}.
This approach has two drawbacks. 
First, it does not take advantage of the nonlinear correlation that may exist between the features of different channels.  
Second, applying the same model to each channel intensifies the computational cost as the number of microphones increases. 

\begin{figure}
    \centering
    \includegraphics[scale=.7]{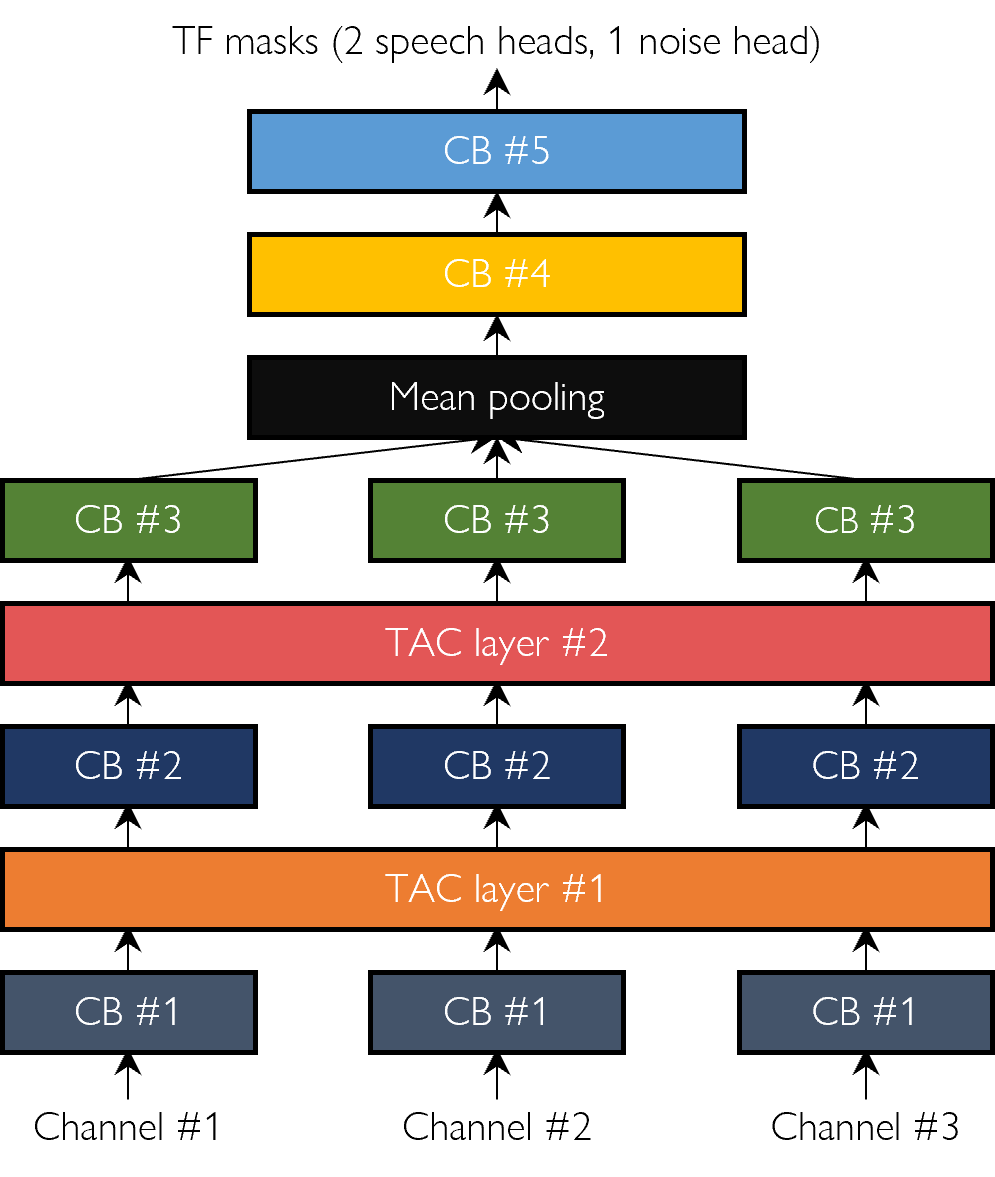}
    \vspace{-.5em}
    \caption{Block diagram of VarArray model. Blocks with the same color share parameters. CB: conformer block.}
    \label{fig: VarArray}
\end{figure}

VarArray was designed to overcome the shortcomings of the SAC
approach. 
Figure \ref{fig: VarArray} shows the block diagram of the model. 
As with SAC, a conformer block is applied to each channel independently at each layer. 
However, the VarArray model has two key differences to address the two problems mentioned above. 

First, TAC layers are interspersed between the conformer blocks. 
TAC aggregates and fuses the features of all input channels in a non-linearly transformed space and feed the outcome back to the individual channels to combine 
the information across the channels efficiently in a permutation-invariant fashion. 
We use a slightly different version of TAC than the original proposal \cite{Luo20-TAC}.
Specifically, denoting the input sequence of the $m$th channel by 
$( \bm{o}_{mt}^{\text{in}} )_{t \in \mathbb{T}}$, 
we compute the output for this channel, $( \bm{o}_{mt}^{\text{out}} )_{t \in \mathbb{T}}$, as follows:
\begin{align}
\bm{o}_{mt}^{\text{out}} = \Biggl[ \text{ReLU} \bigl( A \bm{o}_{mt}^{\text{in}} \bigr), \frac{1}{M} \sum_{\mu \in \mathbb{M}} \text{ReLU} \bigl( B \bm{o}_{\mu t}^{\text{in}} \bigr) \Biggr], 
\label{eq: tac}
\end{align}
where matrices $A$ and $B$ are linear transforms. 

Secondly, 
instead of maintaining the $M$ channels all the way to the last layer, 
we merge them at an earlier layer. 
In the example shown in Fig.~\ref{fig: VarArray}, the model has only one stream
after the third conformer block. 
This significantly reduces the network computation cost.

\subsection{End-to-end optimization}
\vspace{-.3em}

The VarArray model can be further optimized with respect to an ASR-based loss function. 
This is possible with the multi-input multi-output architecture of \cite{chang2019mimo}, which proposed to combine the speech separation TF mask estimator, MVDR beamformer, feature transformation, and  sequence-to-sequence (S2S) ASR model to create an E2E neural network model and perform back-propagation all the way down to the speech separation model. 
We start the E2E optimization with a VarArray model pretrained with simulated multi-channel data. 
During training, we minimize the uPIT-style cross-entropy loss between two predicted hypotheses and reference transcriptions, with the back-end ASR model parameters frozen. 
This allows real transcribed meetings to be used to fine-tune the model parameters and mitigate the mismatch between the simulation and test conditions.


To make the E2E model equivalent to the CSS scheme, 
we add the sliding window stitching process (see Section \ref{sec: css_architecture}), TF mask sparsification, and gain adjustment (see Section \ref{sec: local_separation} for both) to the model. The forward pass processes each multi-channel training sample as follows. 
The training sample is first fed to the VarArray speech separation model to generate the four TF masks described in Section 3.2, and they are sparsified by retaining only the most dominant mask. 
Then, two speech separated signals are computed by using the original multi-channel signal and the MVDR filters derived from the sparse TF masks. 
The generated signals are then chunked into overlapping windows, and the gain adjustment is applied to each windowed signal to equalize the beamformed signal magnitudes with those of the signals that would be obtained by directly applying the TF masks to the input signal. 
The windowed beamformer signals are stitched together to form the two-channel separated signals that are fed to an ASR model. We used the attention-based encoder-decoder model of \cite{wang2021exploring} in our experiments. 
Note that 
these additional elements, especially the gain adjustment, are critical as our training set contains many single-talker samples.

\section{Experimental Results}

We conducted experiments to measure the impact of the proposed array-geometry-agnostic speech separation model on the word error rate (WER) of a meeting transcription system.
Unlike conventional practices of using signal-based metrics or the WER for pre-segmented speech mixtures, 
this allowed us to capture the practical usefulness of the speech separation model. 
Our meeting transcription system, the evaluation tasks, and the results are described below. 

\subsection{System}
\vspace{-.3em}

We employed the audio-visual meeting transcription system of \cite{Yoshioka19d}.
The vision and speaker diarization modules of the original system were removed to focus on the ASR accuracy. 
The processing flow was as follows. 
First, multi-input multi-output
dereverberation was performed in real time with a weighted linear prediction method~\cite{Yoshioka12c}. Then, CSS was applied by using the proposed  speech separation model to generate two output signals that were deemed overlap-free. 
A 1.6-second window was used with a shift of 0.4 seconds, i.e., the stitching was performed by using each 1.2-second overlapping segment. 
Finally, ASR was applied to these signals independently, and the output transcriptions were merged. 
Microsoft's internal system was used for ASR.

The speech separation training set consisted of 438 ($219 \times 2$) hours of speech mixtures that were 
synthesized based on clean speech signals, noise samples, and room impulse responses (RIRs). 
The simulation was carried out by following the recipe of \cite{Yoshioka19d} 
to create 
a balanced mix of the single- and two-speaker training samples as well as the four overlap patterns described in \cite{Yoshioka18}. 
The source speech signals were taken from WSJ SI-284. 
The RIRs were generated by the image method~\cite{Allen79}. 
Both isotropic and directional noise signals were added to the reverberant speech. The isotropic noise samples were synthesized with the algorithm of \cite{Habets07} while the directional noise data were taken from MUSAN~\cite{musan2015}.
The RIRs and the isotropic noise samples were generated for two microphone array geometries, namely AMI and MS (see Section \ref{subsec:task}). For each geometry, 219 hours of audio were generated (which is why the size of the first training set was 438 hours).

The VarArray model was configured as follows. 
As shown in Fig. \ref{fig: VarArray}, our model consisted of conformer blocks and TAC layers. 
Each conformer block comprised five conformer layers, each with four attention heads, 64 dimensions, and 33 convolution kernels. 
There were a total of five conformer blocks, three of which were applied to each stream in parallel. 
For TAC, matrices $A$ and $B$ of Eq. \eqref{eq: tac} halved the number of dimensions for the TAC layer to preserve the dimensionality. 
During training, the number of channels (and the microphones to be used) were randomly picked between 3 and 7 for each mini-batch to expose the model to a variety of array geometries and inter-microphone spacing patterns. 
The training was performed with a mini-batch of 48 four-second sequences. 

\subsection{Tasks}
\label{subsec:task}
\vspace{-.3em}

The evaluation was carried out by using two multi-microphone meeting transcription tasks: AMI (or, more precisely, AMI-MDM) and an internal meetings collection, dubbed as MS. 
For AMI, the ``full-corpus ASR'' partition~\cite{AMI-website} was employed. The audio data were recorded with an eight-channel circular microphone array with a radius of 10 cm. 
Eight-channel and four-channel conditions were considered, where the latter used only the microphones with odd number indices. 
MS is an extension of the test set used in \cite{Yoshioka19d}. 
The same seven-channel microphone array was used. Six microphones were arranged in a circle with a radius of 4.25 cm, and 
the last microphone was located at the center. 
Seven-channel and three-channel conditions were considered, where the latter used the three microphones including the center microphone and constituting a small equilateral triangle. A total of 60 sessions were recorded (150K words in the reference transcriptions). 
The WER was estimated with asclite for AMI while our internal scoring tool was used for the MS test set. 
We used MS to tune system's hyper-parameter values. The AMI development set was actually used as an ``evaluation set''. 

\subsection{Results}
\vspace{-.3em}

\begin{table}[t]
\centering
\caption{\%WERs for AMI-dev and MS. Geometry-dependent models (labeled as Fixed) were trained only for full-array setting of each data set. Training settings 
were optimized for MS.}
\label{tab:result}
\begin{tabu}{l|c|[1pt]ccc|c} \tabucline[1pt]{-}
\multirow{2}{*}{Data} & \multirow{2}{*}{\#Mics} & \multicolumn{3}{c|}{Baselines} & \multirow{2}{*}{VarArray}\\ 
 & & BF1 & SAC & Fixed & \\ \tabucline[1pt]{-}
AMI- & 8 & 25.0 & 18.3 & 18.8 & \textbf{17.7} \\
dev                           & 4 & 25.5 & 18.8  &  --- & \textbf{18.5} \\ \hline
\multirow{2}{*}{MS}  & 7 & 17.6  & 16.0 & 16.0 & \textbf{15.5} \\
                      & 3 & 17.8 & 16.9 & --- & \textbf{15.8} \\ \tabucline[1pt]{-}
\end{tabu}
\vspace{-1em}
\end{table}

Table~\ref{tab:result} shows the WERs of the proposed speech separation method and three baseline systems. 
The first baseline system (BF1) performed super-directive single-output beamforming with real-time beam-steering using our internal beamformer, followed by ASR and thus could not handle speech overlaps. 
This beamformer performed was as effective as BeamformIt~\cite{Anguera07} in our internal test. 
The second baseline system was based on SAC and  
applied a single-stream separation model to each input stream of Fig. \ref{fig: VarArray} with the magnitude and IPD pair as input to obtain TF masks from each stream.
As with \cite{Wang19-spectralspatial}, the output speech orders were aligned across the streams, and then the TF masks were averaged to be used for beamforming. 
This separation model had the same number of conformer blocks as our proposed VarArray system. 
It was trained on the same data set as ours by randomly choosing the input stream. 
The third baseline system used a concatenation of the magnitude spectra of the first microphone and the IPDs between the first microphone and the rest of the microphones. The model was trained for AMI and MS separately based on the corresponding subsets of the training set due to the input dependency on the number of microphones. 

For all the four conditions, our system significantly outperformed the second baseline system, showing VarArray's capability of effectively leveraging the spatial information for different microphone array geometries. 
The proposed model also outperformed the array-geometry-dependent models for both test sets. This could be largely attributed to the fact that the array-geometry-agnostic model was able to take advantage a larger training set.  

\begin{table}[t]
\centering
\caption{Performance dependency on microphone array configuration of training set.}
\vspace{.3em}
\label{tab:trainset}
\begin{tabu}{l|[1pt]c|c|c} \tabucline[1pt]{-}
\diagbox[width=5em]{Test}{Train}&AMI & MS  & AMI+MS\\ \tabucline[1pt]{-}
 AMI-dev & 18.0 & 18.8 & \textbf{17.7} \\
 MS & 15.7 & 15.8 & \textbf{15.5} \\ \tabucline[1pt]{-}
\end{tabu}
\vspace{-1em}
\end{table}

Two additional VarArray models were further trained to examine
the performance dependency on the array configuration of the training set. One model was trained on the subset that was based on the AMI geometry RIRs. The other model used the MS-geometry portion of the training set.
Table~\ref{tab:trainset} shows the experimental results. 
It was observed that the separation model trained with both microphone array configurations performed the best for both test sets. Nonetheless, the other two models also showed good generalization ability to unseen array geometries. 
It is noteworthy that the model trained on the AMI-geometry subset performed slightly better than the one trained on the MS-geometry subset. We presume that this is because the AMI-geometry portion of the training set had greater diversity in terms of the inter-microphone distances.



\begin{table}[t]
\centering
\caption{Impact of end-to-end optimization on \%WER.}
\vspace{.3em}
\label{tab:e2eoptimazation}
\begin{tabu}{c|c|cc|cc} \tabucline[1pt]{-}
\multirow{2}{*}{Pre-Train} & {E2E} & \multicolumn{2}{c|}{AMI-dev}  & \multicolumn{2}{c}{AMI-eval} \\ 
 & {optim} & w/o ovlp & w/ ovlp & w/o ovlp & w/ ovlp \\ \hline
MS &  & 15.3 & 18.7 & 17.4 & 22.2   \\
MS & \checkmark & {\bf 14.8} & {\bf 18.2} & {\bf 16.8} & {\bf 21.5} \\ 
AMI+MS &  & 15.5 & 17.8 & 17.0 & 20.5   \\
AMI+MS & \checkmark & {\bf 14.7} & {\bf 17.5} & {\bf 16.8} & {\bf 20.4} \\ 
\tabucline[1pt]{-}
\end{tabu}
\vspace{-1em}
\end{table}

Table~\ref{tab:e2eoptimazation} shows the E2E optimization experiment results for the AMI development and evaluation sets. We segmented the original long-form AMI training  recordings by silence positions, removed the segments shorter than 10 seconds, and picked only the segments consisting of one or two speakers, which resulted in a 8.52-hour training set. 
We experimented two seed models: one trained on the MS-geometry data, one on the AMI- and MS-geometry data. Note that, in this experiment, we used a larger pool of clean speech than in the previous experiments, including samples from our internal data,  while we used the same RIR and noise data. The training set size for the seed models was 1,500 hours for the MS- and AMI-geometry portions each. Therefore, the WER numbers are slightly different from those of the previous experiments. 
Consistent WER gains were observed, suggesting the E2E optimization using the small amount of real transcribed data to help mitigate the mismatch between the training and test conditions. 
Larger gains were observed for the model pre-trained on the MS-geometry data, indicating that the E2E optimization addressed both the geometry difference and the mismatch between the simulated and real data. It is also noteworthy that the gains were more prominent for non-overlap regions. This could mean that the E2E fine-tuning made the separated signals more friendly to ASR rather than removing interfering signals. 


\section{Conclusion}

We described VarArray, an array-geometry-agnostic speech separation model. Extensive evaluation was conducted with a CSS framework in two real meeting transcription tasks. 
The proposed model outperformed the SAC approach by efficiently leveraging the spatial information. It was also shown that the VarArray model could generalize to unseen array geometries  well and that the E2E optimization could mitigate the mismatch between the training and test conditions. 
The array-geometry-agnostic modeling is useful for production. In parallel to this work, we examined its impact on personalized noise reduction \cite{Taherian22}. Further investigation in different tasks is desired.

\vfill\pagebreak

\bibliographystyle{IEEEbib}
\bibliography{my_references,refs}

\end{document}